\documentclass{article}
\usepackage{times}
\usepackage{amsfonts}
\usepackage{graphicx}
\begin{document}
\hyphenation{Min-kows-kian}
\hyphenation{pre-sent-ly}

\noindent
{\Large  SPACETIME METRIC FROM QUANTUM--GRAVITY \\CORRECTED FEYNMAN PROPAGATORS}
\vskip1cm
\noindent
{\bf P. Fern\'andez de C\'ordoba}$^{1}$, {\bf J.M. Isidro}$^{2}$ and {\bf Rudranil Roy}$^{3}$\\
Instituto Universitario de Matem\'atica Pura y Aplicada,\\ Universitat Polit\`ecnica de Val\`encia, Valencia 46022, Spain\\
$^{1}${\tt pfernandez@mat.upv.es}, $^{2}${\tt joissan@mat.upv.es},\\
$^{3}${\tt rudranilroy.gravity@gmail.com}
\vskip.5cm
\noindent

\noindent
{\bf Abstract}  Differentiation of the scalar Feynman propagator with respect to the spacetime coordinates yields the metric on the background spacetime that the scalar particle propagates in. Now Feynman propagators can be modified in order to include quantum--gravity corrections as induced by a zero--point length $L>0$. These corrections cause the length element $\sqrt{s^2}$ to be replaced with $\sqrt{s^2 + 4L^2}$ within the Feynman propagator. In this paper we compute the metrics derived from both the quantum--gravity free propagators and from their quantum--gravity corrected counterparts. We verify that the latter propagators yield the same spacetime metrics as the former, provided one measures distances greater than the quantum of length $L$. We perform this analysis in the case of the background spacetime $\mathbb{R}^D$ in the Euclidean sector.

\tableofcontents

\section{Introduction}\label{anfang}

The search for a theory of quantum gravity  is an ongoing effort that has kept theoreticians busy for almost a century by now \cite{KIEFER1}. While a full--fledged theory of quantum gravity still remains elusive (see, {\it e.g.}\/,  ref. \cite{KIEFER2} for a recent review), under certain assumptions there are some quantum--gravity corrections that can be more or less readily computed. For example, it is generally believed that the existence of a quantum of length $L>0$ must be a universal feature of quantum gravity in the mesoscopic regime \cite{CASADIO1, CASADIO2, FINSTER, GARAY, KOTHAWALA1, KOTHAWALA2, KOTHAWALA3, PADDY0, PADDY1, PADDY3, PADDY2, SINGH}. 

Endowing a $D$--dimensional manifold ${\cal M}$ with a Riemannian metric $g$,
\begin{equation}
{\rm d}s^2=g_{ij}(x){\rm d}x^{i}{\rm d}x^j,
\label{metrik}
\end{equation}
permits one to consider the classical motion of a particle on the spacetime ${\cal M}$. For example a free, massive scalar has an action integral defined as (the mass $m$ times) the proper length, $mS(x,y)=m\int_y^x{\rm d}s$. The corresponding quantum particle is governed by the Feynman propagator
\begin{equation}
G_D(x,y)=\sum_{\rm paths}\exp\left[-mS(x,y)\right].
\label{treintayuno}
\end{equation}
The propagator $G_D(x,y)$ depends on the points $x,y$ through the square of the geodesic distance $s^2$ between $x,y$. Let $G_D=f(s^2)$ formally denote the functional expression for the Feynman propagator as a certain function of $s^2$ (see, {\it e.g.}\/, Eq. (\ref{catorce}) below). Assume that one can invert it in order to solve for $s^2$ as a function of the propagator, and let us express this inverse relation as $s^2=f^{-1}(G_D)$. Since by Eq. (\ref{metrik}) we have
\begin{equation}
g_{ij}(x)=\frac{\partial^2 }{\partial x^{i}\partial x^j}s^2(x),
\label{teile}
\end{equation}
we can formally write
\begin{equation}
g_{ij}(x)=\colon\lim_{y\to x}\frac{\partial^2 }{\partial x^{i}\partial y^j}f^{-1}\left(G_D(x,y)\right)\colon.
\label{metprop}
\end{equation}
Above, provision has been made for the fact that the Feynman propagator $G_D(x,y)$ is singular when $x=y$. So one first splits the points $x$ and $y$ a short distance apart, next one differentiates with respect to $x$ and $y$, finally one lets $y\to x$; a procedure which is reminiscent of certain regularisation prescriptions in quantum field theory. The colons :: in Eq. (\ref{metprop}) remind us that some prescription must be adopted in order to extract a finite answer from the singularity. We will explain our prescription in due time.

Although in principle the procedure just laid out allows one to derive the metric $g_{ij}$ from a knowledge of the Feynman propagator $G_D(x,y)$, in practice there is little hope of ever being able to compute the inverse function $s^2=f^{-1}(G_D)$ explicitly. This notwithstanding, some prescriptions to derive geometric objects from quantum--mechanical quantities have been proposed in the literature \cite{MARTINMARTINEZ, SARAVANI}, that circumvent the computational difficulty of Eq. (\ref{metprop}). One of the aims of this paper is to put forward an alternative prescription that, while computationally tractable, yields back the correct spacetime metric.

We first define the {\it  Feynman bitensor}\/ $F_{ij}(x,y)$ \cite{KOTHAWALA}
\begin{equation}
F_{ij}(x,y)= N_D\frac{\partial^2 }{\partial x^{i}\partial y^j}G_D(x,y),
\label{abelkabron}
\end{equation}
where $N_D$ is some dimension--dependent normalisation factor, to be determined presently. Now $F_{ij}(x,y)$ qualifies as a bitensor because the Feynman propagator $G_D(x,y)$ is a biscalar, hence its derivatives transform as a bitensor. In the coincidence limit $y\to x$ one defines the {\it  Feynman tensor}\/ $F_{ij}(x)$
\begin{equation}
F_{ij}(x)=N_D\lim_{y\to x}\frac{\partial^2 }{\partial x^{i}\partial y^j}G_D(x,y).
\label{abelcabron}
\end{equation}
However, as $y\to x$, the right--hand side of (\ref{abelkabron}) will contain singular terms in a Laurent series expansion in powers of $s$, thus rendering (\ref{abelcabron}) ill defined.  Assuming some regularisation has been adopted, the sought--for metric will then be
\begin{equation}
g_{ij}(x)=\colon F_{ij}(x)\colon=N_D\colon\lim_{y\to x}\frac{\partial^2 }{\partial x^{i}\partial y^j}G_D(x,y)\colon
\label{calon}
\end{equation}

We see that, modulo irrelevant normalisation factors,  $g_{ij}(x)$ in Eq. (\ref{calon}) differs from $g_{ij}(x)$ in Eq. (\ref{metprop}) by the absence of the inverse function $f^{-1}$. To arrive at Eq. (\ref{calon}) we differentiate the Feynman propagator twice, take the coincidence limit, and extract some finite term from the singular terms in a Laurent expansion in $s$.  We will see that {\it the metric $g_{ij}$ of the underlying spacetime manifold ${\cal M}$ will turn out to be encoded in the coefficients multiplying those singularities}\/. The operation of taking the colons :: will consist in judiciously picking out the coefficient of an appropriate singular term on the right--hand side of Eq. (\ref{abelcabron}). 

Eq. (\ref{calon}) correctly defines a symmetric, rank 2 tensor on ${\cal M}$. That it actually defines a Riemannian metric requires proving that it is also positive definite. This latter point will be established once one specifies how the operation of taking the colons :: actually works, {\it i.e.}\/, what particular coefficient in the Laurent expansion is picked. One can then {\it a posteriori}\/ verify that the chosen term in the Laurent expansion does indeed define a positive--definite Riemannian metric.

We can push the previous arguments a bit further and consider modifications to the standard Feynman propagator $G_D$ on ${\cal M}$. By this we mean objects that will be constructed similarly to $G_D$, but with some added requirements. For example, one might require of $G_D$ that it be invariant under the duality transformation \cite{PADDY0, PADDY1, PADDY3, PADDY2}
\begin{equation}
S\longrightarrow\frac{L^2}{S}.
\label{sesenta}
\end{equation}
Above, $L$ is a {\it quantum of length}\/ on the spacetime ${\cal M}$. Imposing the principle of invariance under the duality (\ref{sesenta}) is one way of implementing ultraviolet (UV) completion \cite{CROWTHER, ISIDRO, NOI} in an eventual theory of quantum gravity. In our case this is achieved as follows. One modifies the standard path integral (\ref{treintayuno}) into the manifestly duality--invariant path integral
\begin{equation}
G_D^{{\rm (QG)}}(x,y)=\sum_{\rm paths}\exp\left\{-m\left[S(x,y)+\frac{L^2}{S(x,y)}\right]\right\}.
\label{treintaydos}
\end{equation}
The superscript QG in the modified Feynman propagator (\ref{treintaydos}) stands for {\it quantum gravity}\/. It has been established in refs. \cite{PADDY0, PADDY1, PADDY3, PADDY2} that, in the particular case of a massive scalar particle on ${\cal M}$, the propagator (\ref{treintaydos}) includes lowest--order, quantum--gravitational effects due to the presence of  the zero--point  length $L$. 

We can now substitute the modified propagator (\ref{treintaydos}) into Eq. (\ref{abelkabron}) to define a quantum--gravity corrected Feynman bitensor $F^{\rm (QG)}_{ij}(x,y)$
\begin{equation}
F^{\rm (QG)}_{ij}(x,y)= N^{\rm (QG)}_D\frac{\partial^2 }{\partial x^{i}\partial y^j}G_D^{{\rm (QG)}}(x,y),
\label{20}
\end{equation}
as well as the corresponding Feynman tensor  $F^{\rm (QG)}_{ij}(x)$:
\begin{equation}
F^{\rm (QG)}_{ij}(x)= N^{\rm (QG)}_D\lim_{y\to x}\frac{\partial^2 }{\partial x^{i}\partial y^j}G_D^{{\rm (QG)}}(x,y).
\label{nikil}
\end{equation}
{}From here one derives the quantum--gravity corrected metric
\begin{equation}
g^{\rm (QG)}_{ij}(x)=\colon F^{\rm (QG)}_{ij}(x)\colon=N_D^{\rm (QG)}\colon\lim_{y\to x}\frac{\partial^2 }{\partial x^{i}\partial y^j}G_D^{\rm (QG)}(x,y)\colon
\label{2}
\end{equation}
Although there is no {\it a priori}\/ reason for $g^{\rm (QG)}_{ij}$ to coincide with its parent metric $g_{ij}$, they will actually be the same {\it provided one measures distances greater than the quantum of length $L$}. The reason is the form invariance of the Feynman propagators before and after imposing invariance under the duality transformation (\ref{sesenta}) (see Eqs. (\ref{catorce}), (\ref{kagonesteve}) below).  

In this work we first consider the known  Feynman propagator for a free, massive scalar particle on the Euclidean spacetime manifold $\mathbb{R}^D$ and validate our prescription (\ref{calon}) by verifying that the original metric is retrieved as expected. Next we move on to the duality--invariant Feynman propagator derived in ref. \cite{PADDY0, PADDY1, PADDY2}, compute the corresponding metric via our prescription (\ref{2}) and verify that, for distances larger than $L$, it coincides with that obtained from the original ({\it i.e.}\/, non duality invariant) Feynman propagator.

For simplicity we will limit our analysis to the flat spacetime ${\cal M}=\mathbb{R}^D$ endowed with the standard Euclidean metric; in upcoming work we will analyse curved spacetimes \cite{UPCOMING}. In our use of special functions we follow the conventions of ref. \cite{GRADSHTEYN}.

\section{The metric on $\mathbb{R}^D$ from Feynman propagators}

\subsection{Propagators}

A massive, free scalar particle in Euclidean space $\mathbb{R}^D$ has the Feynman propagator 
\begin{equation}
G_{D}(s)=\frac{m^{D-2}}{(2\pi)^{D/2}}\frac{K_{D/2-1}\left(ms\right)}{\left(ms\right)^{D/2-1}}, \qquad s^2(x,y)=\sum_{i=1}^D(x^i-y^i)^2,
\label{catorce}
\end{equation}
where $K_n(z)$ is a modified Bessel function and $s^2(x,y)$ is the square of the geodesic distance between points $x,y\in\mathbb{R}^D$. By homogeneity and isotropy, $x,y$ enter the Feynman propagator only through the  combination $s$. 

Corresponding to the gravity--free propagator (\ref{catorce}) we have a duality--invariant propagator \cite{PADDY0, PADDY1, PADDY2}
\begin{equation}
G^{\rm (QG)}_D(s)=\frac{m^{D-2}}{(2\pi)^{D/2}}\frac{K_{D/2-1}\left(m\sqrt{s^2+4L^2}\right)}{\left(m\sqrt{s^2+4L^2}\right)^{D/2-1}}.
\label{kagonesteve}
\end{equation}
We observe that the functional form of the Feynman propagator is the same in Eqs. (\ref{catorce}) and (\ref{kagonesteve}), the only difference being the shift 
\begin{equation}
s=\sqrt{s^2}\longrightarrow\sqrt{s^2+4L^2}
\label{chif}
\end{equation}
due to the zero--point length $L>0$.

\subsection{Summary of results}

In this section we summarise our main results; computational details are presented in the appendix.

\subsubsection{$L=0$}

In the absence of a quantum of length one arrives at a Feynman bitensor (see Eqs. (\ref{steveabelmarikon}), (\ref{scheissesteve}) and (\ref{steveabelbobo}))
\begin{equation}
F_{ij}(x,y)=N_D\frac{\partial^2}{\partial x^i\partial y^j}G_D(x,y)=\frac{1}{s^D}\left[\delta_{ij}-\frac{D}{s^2}\left(x^i-y^i\right)\left(x^j-y^j\right)\right],
\label{1}
\end{equation}
and the normalisation factor turns out to be
\begin{equation}
N_{D}=\frac{2\pi^{D/2}}{\Gamma(D/2)}.
\label{3}
\end{equation}
In this case the regularisation prescription (\ref{calon}) reads: {\it the metric is given by the coefficient multiplying the least singular term in the Laurent expansion (\ref{1})}\/. That is,
\begin{equation}
g_{ij}(x)=\colon F_{ij}(x)\colon=\delta_{ij}.
\label{4} 
\end{equation}

\subsubsection{$L>0$}

Here the Feynman bitensor (\ref{1}) gets replaced with (see Eqs. (\ref{steveabelkaka}), (\ref{stevescheisse}) and (\ref{japili}))
\begin{equation}
F^{\rm (QG)}_{ij}(x,y)=N^{\rm (QG)}_D\frac{\partial^2}{\partial x^i\partial y^j}G^{\rm (QG)}_D(x,y)
\label{5}
\end{equation}
$$
=\frac{1}{\left(\sqrt{s^2+4L^2}\right)^D}\left[\delta_{ij}-\frac{D}{s^2+4L^2}\left(x^i-y^i\right)\left(x^j-y^j\right)\right],
$$
while the normalisation factor remains as above:
\begin{equation}
N^{\rm (QG)}_{D}=\frac{2\pi^{D/2}}{\Gamma(D/2)}.
\label{6}
\end{equation}
The expansion (\ref{5}) is no longer singular at $s=0$ because the zero--point length $L>0$ prevents it.  Without loss of generality we can set $x=0$ and consider the coincidence limit to be $y\to 0$:
\begin{equation}
\lim_{y\to 0}F^{\rm (QG)}_{ij}(y)=\lim_{y\to 0}\left[\frac{1}{\left(2L\right)^D}\left(\delta_{ij}-\frac{D}{4L^2}y^iy^j\right)\right].
\label{7}
\end{equation}
However, we cannot simply set $\lim_{y\to 0}y^iy^j=0$, because that would amount to having $L=0$. Regularisation now stands for the following prescription: if $\vert\vert{\bf y}\vert\vert<L$ we take
\begin{equation}
: \lim_{y\to 0}F^{\rm (QG)}_{ij}(y):=\delta_{ij}-\frac{D}{4L^2}y^iy^j,
\label{11}
\end{equation}
while, if $\vert\vert{\bf y}\vert\vert\geq L$, we take
\begin{equation}
: \lim_{y\to 0}F^{\rm (QG)}_{ij}(y):=\delta_{ij}.
\label{12}
\end{equation}
The overall factor $(2L)^{-D}$ plays no role and has been cancelled. Altogether, the metric $g_{ij}^{\rm (QG)}$ reads
\begin{equation}
g_{ij}^{\rm (QG)}(y)=\left\{\begin{array}{ll}
\delta_{ij}-\frac{D}{4L^2}y^iy^j,\quad&\vert\vert{\bf y}\vert\vert<L\\
\delta_{ij},\quad &\vert\vert{\bf y}\vert\vert \geq L.
\end{array}\right.
\label{15}
\end{equation}

\subsection{Discussion}

Spacetime cannot be probed at distances shorter than $L$. Hence the term $-(D/4)y^iy^j$ correcting $\delta_{ij}$ in (\ref{15}) is to be understood as {\it implementing a change in the metric}\/ across the interface $\vert\vert{\bf y}\vert\vert=L$. The interior of the sphere $\vert\vert{\bf y}\vert\vert=L$ still makes sense as a mathematical construct, but not as a description of physical reality {\it as seen by an observer sitting inside}\/. To reiterate: {\it there is no physical metric for values of ${\bf y}$ such that $\vert\vert{\bf y}\vert\vert<L$}. The QG--corrected metric (\ref{11}) must be regarded as being merely the coefficient multiplying the relevant term in a Laurent expansion. The QG--corrected Feynman propagator imposes a natural cutoff that renders the notion of a metric inside the sphere $\vert\vert{\bf y}\vert\vert=L$ physically meningless.

The transition referred to above may, but need not, be a discontinuity such as that in Eq. (\ref{15}). In fact such a discontinuity could be removed altogether and replaced with a continuous, even a differentiable transition across the sphere $\vert\vert{\bf y}\vert\vert=L$, smoothly distributed over a few fractions of $L$ in length. The key property is the existence of a transition (be it continuous or not, smooth or not) as one traverses the interface between the two regimes; this transition prevents an observer from accessing any information carrying momenta larger than $1/L$.

Our prescription correctly captures the fact that, far enough from the origin, the quantum of length $L>0$ becomes unobservable. Then the coarse--grained description of the metric structure on $\mathbb{R}^D$ remains as given by the standard Euclidean metric $\delta_{ij}$.

The above--mentioned transition is localised not just  on a sphere around the origin ${\bf y}={\bf 0}$. Rather, it occurs around every single point ${\bf y}\in\mathbb{R}^D$, once one zooms in down to lengths of order $L$ from ${\bf y}$. Of course, by homogeneity of $\mathbb{R}^D$ we can limit our discussion to the point ${\bf y}={\bf 0}$. 

While the line element (\ref{metrik}) qualifies as a {\it bona fide}\/ Riemannian metric, the inhomogeneous term $4L^2$ prevents one from interpreting $s^2+4L^2$ as a quadratic form, hence also as a metric in the strict sense of the word. On the other hand, the QG--corrected metric (\ref{15}) {\it does}\/ retain its interpretation as a quadratic form.

\section{Appendix: computation of the Feynman bitensor}

Here we compute the Feynman bitensor (\ref{abelkabron}) and its quantum--gravity corrected counterpart (\ref{20}), in order to identify the corresponding metrics. We will work in increasing values of the dimension $D$, separating the cases $L=0$ and $L>0$.

{}For use in connection with Eqs. (\ref{catorce}) and (\ref{kagonesteve}) we recall the expansions
\begin{equation}
K_0(z)=-\frac{1}{2}\sum_{l=0}^{\infty}\frac{1}{(l!)^2}\left(\frac{z}{2}\right)^{2l}\left[2\ln\left(\frac{z}{2}\right)-2\psi(l+1)\right]
\label{kool}
\end{equation}
where $\psi(z)$ is the digamma function, and
\begin{equation}
K_{n}(z)\simeq\frac{(n-1)!}{2}\left(\frac{2}{z}\right)^n, \quad z\to 0,\quad n=1,2,\ldots
\label{stevekaka}
\end{equation}
When the order $n$ of the Bessel function equals an integer plus one half we have 
\begin{equation}
K_{n+1/2}(z)=\sqrt{\frac{\pi}{2z}}\,{\rm e}^{-z}\,\sum_{l=0}^n\frac{(n+l)!}{l!(n-l)!(2z)^l}.
\label{merdeabel}
\end{equation}
Use of the previous expansions is justified because we are interested in the behaviour of the Feynman propagator in a neighbourhood of the origin, where it becomes singular. We will therefore keep just the leading singularity as $z\to 0$.

\subsection{$D=2$ with $L=0$}

The Feynman propagator in a neighbourhood of the origin reads, by Eqs. (\ref{catorce}) and (\ref{kool}),
\begin{equation}
G_2(s)\simeq -\frac{1}{2\pi}\ln \left(\frac{m\,{\rm e}^{\gamma}}{2}s\right), \qquad s\to 0,
\label{gedos}
\end{equation}
where $\gamma=-\psi(1)$ is the Euler--Mascheroni constant. A short computation produces the Feynman bitensor
\begin{equation}
2\pi\frac{\partial}{\partial x^i}\frac{\partial}{\partial y^j}G_2(s)=\frac{1}{s^2}\left[\delta_{ij}-\frac{2}{s^{2}}(x^i-y^i)(x^j-y^j)\right].
\label{steveabelmarikon}
\end{equation}

\subsection{$D=2$ with $L>0$}

Next we let $L>0$. By Eqs. (\ref{kagonesteve}) and (\ref{kool}), in a neighbourhood of the origin the Feynman propagator reads
\begin{equation}
G_2^{\rm (QG)}(s)\simeq -\frac{1}{2\pi}\ln\left(\frac{m\,{\rm e}^{\gamma}}{2}\sqrt{s^2+4L^2}\right), \qquad s\to 0,
\label{pera}
\end{equation}
and for the Feynman bitensor one finds
\begin{equation}
2\pi\frac{\partial}{\partial x^i}\frac{\partial}{\partial y^j}G_2^{\rm (QG)}(s)=\frac{1}{\left(\sqrt{s^2+4L^2}\right)^2}\left[\delta_{ij}-\frac{2}{s^2+4L^2}(x^i-y^i)(x^j-y^j)\right].
\label{steveabelkaka}
\end{equation}
The above correctly reduces to the Feynman bitensor (\ref{steveabelmarikon}) when $L=0$. We observe that, as expected, the Laurent expansion (\ref{steveabelkaka}) is no longer in powers of $s^2$ but in powers of its shifted counterpart $s^2+4L^2$ instead. This is consistent with the fact, mentioned in connection with Eq. (\ref{kagonesteve}), that the functional dependence (on the line element) of the original and of the corrected propagator is the same in both cases.

\subsection{$D=2k$ with $k>1$ and $L=0$}

By Eq. (\ref{stevekaka}), the Feynman propagator (\ref{catorce}) becomes, in a neighbourhood of the origin,
\begin{equation}
G_{2k}(s)\simeq\frac{(k-2)!}{4\pi^k}s^{2-2k}, \qquad s\to 0.
\label{abelbaboso}
\end{equation}
{}For the Feynman bitensor (\ref{abelcabron}) one finds
\begin{equation}
\frac{2\pi^k}{(k-1)!}\frac{\partial}{\partial x^i}\frac{\partial}{\partial y^j}G_{2k}(s)=\frac{1}{s^{2k}}\left[\delta_{ij}-\frac{2k}{s^2}\left(x^i-y^i\right)\left(x^j-y^j\right)\right].
\label{scheissesteve}
\end{equation}
The above results correctly reproduce those of $D=2$ upon setting $k=1$, even if the Bessel functions involved in the cases $k=1$ and $k>1$ have different behaviours around the origin. For this reason they have been treated separately.

\subsection{$D=2k$ with $k>1$ and $L>0$}

As usual we need the behaviour of the quantum--gravity corrected Feynman propagator (\ref{kagonesteve}) in a neighbourhood of the origin: by (\ref{stevekaka}), this is
\begin{equation}
G^{\rm (QG)}_{2k}(s)\simeq\frac{(k-2)!}{4\pi^k}(s^2+4L^2)^{1-k}, \qquad s\to 0.
\label{abelburro}
\end{equation}
Using this one computes the Feynman bitensor
\begin{equation}
\frac{2\pi^k}{(k-1)!}\frac{\partial}{\partial x^i}\frac{\partial}{\partial y^j}G_{2k}^{\rm (QG)}(s)
\label{stevescheisse}
\end{equation}
$$
=\frac{1}{(\sqrt{s^2+4L^2})^{2k}}\left[\delta_{ij}-\frac{2k}{s^2+4L^2}\left(x^i-y^i\right)\left(x^j-y^j\right)\right],
$$
which reduces to that in Eq. (\ref{scheissesteve}) upon setting $L=0$.

\subsection{$D=2k+1$ with $k\geq1$ and $L=0$}

In this case we have, in a neighbourhood of the origin,
\begin{equation}
G_{2k+1}(s)\simeq \frac{1}{(4\pi)^k}\frac{(2k-2)!}{(k-1)!}\,s^{1-2k}, \qquad s\to 0,
\label{8}
\end{equation}
and therefore
\begin{equation}
(4\pi)^k\frac{(k-1)!}{(2k-1)!}\frac{\partial}{\partial x^i}\frac{\partial}{\partial y^j}G_{2k+1}(s)=\frac{1}{s^{2k+1}}\left[\delta_{ij}-\frac{2k+1}{s^2}(x^i-y^i)(x^j-y^j)\right].
\label{steveabelbobo}
\end{equation}

\subsection{$D=2k+1$ with $k\geq1$ and $L>0$}

In a neighbourhood of the origin the Feynman propagator reads
\begin{equation}
G_{2k+1}^{\rm (QG)}(s)\simeq\frac{1}{(4\pi)^k}\frac{(2k-2)!}{(k-1)!}\,\left(s^2+4L^2\right)^{-k+1/2}, \quad s\to 0.
\label{10}
\end{equation}
Proceeding as above one arrives at the Feynman bitensor
\begin{equation}
(4\pi)^k\frac{(k-1)!}{(2k-1)!}\frac{\partial}{\partial x^i}\frac{\partial}{\partial y^j}G_{2k+1}^{\rm (QG)}(s)
\label{japili}
\end{equation}
$$
=\frac{1}{\left(\sqrt{s^2+4L^2}\right)^{2k+1}}\left[\delta_{ij}-\frac{2k+1}{s^2+4L^2}(x^{i}-y^{i})(x^j-y^j)\right],
$$
which reduces to that in Eq. (\ref{steveabelbobo}) upon setting $L=0$.

\vskip1cm
\noindent
{\bf Acknowledgments} Work funded by FEDER/MCIN under grant PID2021-128676OB-I00 (Spain).

\end{document}